\documentclass[11pt,moriond]{article}
\usepackage{moriond}
\usepackage[pdftex]{graphicx}
\usepackage{amsmath}
\usepackage{hyperref}
\def\PRD{{\em Phys. Rev.} D}

\def\jgr{\em Journal of Geophysical Research }
\def\nar{\em New Astr. Rev }
\def\nat{\em Nature }
\def\mnras{\em MNRAS }
\def \o2{$O_{2}$}

\def\be{\begin{equation}}
\def\ee{\end{equation}}
\def\bea{\begin{eqnarray}}
\def\eea{\end{eqnarray}}

\begin{document}

%\title{A template of atmospheric $\mathbf{O_{2}}$ circularly polarized emission for CMB experiments}
\title{A template of atmospheric molecular oxygen circularly polarized emission for CMB experiments}

\author{ G. Fabbian$^{1}$, S. Spinelli$^{2}$, M. Gervasi$^{2}$, A. Tartari$^{1,2}$, M. Zannoni$^{2}$}

\address{$^{1}$AstroParticule et Cosmologie, Univ Paris Diderot, CNRS/IN2P3, CEA/Irfu, Obs de Paris, Sorbonne Paris Cit\'e, France \\$^{2}$ Dipartimento di Fisica ``G. Occhialini'', Universit\`a di Milano Bicocca, Piazza della Scienza 3, 20126, Milano}

\maketitle\abstracts{
We compute the polarized signal from atmospheric molecular oxygen due to Zeeman effect in the Earth magnetic field for various sites suitable for CMB measurements such as South Pole, Dome C (Antarctica) and Atacama desert (Chile). We present maps of this signal for those sites and show their typical elevation and azimuth dependencies. We find a typical circularly polarized signal (V Stokes parameter) level of $50 - 300$ $\mu K$ at 90 GHz when looking at the zenith; Atacama site shows the lowest emission while Dome C site presents the lowest gradient in polarized brightness temperature (0.3 $\mu K/ \circ$ at 90 GHz). The accuracy and robustness of the template are tested with respect to actual knowledge of the Earth magnetic field, its variability and atmospheric parameters.}
\section{Motivation}
In the last 15 years several experiments managed to successfully observe CMB anisotropies from the ground inside the atmospheric windows far from the emission lines of most abundant molecules of the atmosphere. For the next generation of  ground-based experiments, which will be able to map CMB polarization at high resolution and with high sensitivity, an evaluation of atmospheric polarized emission is required. As pointed out first by Hanany {\it et al.}\cite{hanany}, the main atmospheric contaminant for this kind of measurements is the circularly polarized Zeeman emission of molecular oxygen ($O_{2}$). Such signal can in fact be converted into linear polarization through non idealities of the instrument and thus generate spurious Q and U quantities.\\*
In second instance the circular polarization of CMB itself has a cosmological interest and a theoretical effort is undergoing to investigate those mechanisms that can produce such signal either at last scattering surface or at later epoch (see for example Giovannini,\cite{giovannini} Cooray {\it et al.}\cite{cooray} and references in Spinelli {\it et al.}\cite{spinelli}). In this perspective any dedicated future measurement from the ground aiming to update the current constraint on V Stokes parameter \cite{partridge} has to face the presence of atmospheric molecular oxygen as a foreground.
\section{Polarized emission theory}
\o2 is the only abundant molecule in the atmosphere having a non-negligible magnetic dipole moment since its electronic configuration sets two electrons coupled with parallel spin in the highest energy level. This characteristic, together with the presence of the Earth magnetic field, allows the Zeeman splitting of its roto-vibrational lines in the millimeter region of the electromagnetic spectrum. In particular we notice that \o2 has a single intense line around $ \nu \simeq 118.75 $ GHz and a forest of lines in the range $50-70$ GHz. Selection rules corresponding to transitions $\Delta j=\pm 1$ and $\Delta m_{j}=0,\pm 1$ identify three different type of Zeeman lines ($\pi$, $\sigma_{\pm}$ respectively) having polarization direction and intensity properties varying as a function of the angle $\theta$ between the line of sight (los) and the Earth magnetic field.\\
%The polarized intensity for each line can be evaluated through their coherency matrices ${\bf\rho}$; the total coherency matrix $\mathbf{A}_{tot}$ is given by the superposition of those for all the different lines weighted by two factors taking into account the transition dependence on pressure and temperature ($C(\nu, P, T))$ and the line broadening and mixing ($F(\nu, \nu_{k}, \Delta\nu_{c}) )$.\cite{rosenkranz}
The polarized intensity of each line can be evaluated through its coherency matrix ${\bf\rho}$; the total coherency matrix $\mathbf{A}_{tot}$ is given by the superposition of those for all the different lines weighted by two factors taking into account the transition dependence on pressure and temperature ($C(\nu, P, T))$ and the line broadening and mixing ($F(\nu, \nu_{k}, \Delta\nu_{c}) )$.\cite{rosenkranz}
\be
\mathbf{A}_{tot}=C(\nu,P,T)\sum_{\Delta m_{j}=-1}^{+1}\mathbf{\rho}_{\Delta m_{j}}\sum_{m_{j}=-j}^{+j}P_{trans}(S,L,m_{j},\Delta j,\Delta m_{j})F(\nu,\nu_{k},\Delta\nu_{c})
\label{atotsum}
\ee
\noindent
$P_{trans}$ denotes the transition probability for each line as predicted by quantum mechanics and fixes their relative intensities. Taking into account all these relations, the final matrix has the following form:
\be
\mathbf{A}_{tot}=a\mathbf{\rho}_{\sigma_{-}}+b\mathbf{\rho}_{\sigma_{+}}+c\mathbf{\rho}_{\pi}=\begin{pmatrix} a+b& i(a-b)\cos\theta\\-i(a-b)\cos\theta& (a+b)\cos^{2}\theta+c\sin^{2}\theta\\\end{pmatrix}
\ee
\noindent
We notice that the presence of non vanishing, conjugate and purely imaginary off-diagonal terms means that radiation is only circularly polarized ($V\neq0, U=0$) with a circular polarization stronger when the los is aligned with the Earth magnetic field ($V\propto \cos\theta$). On the other hand, diagonal terms are different and thus a small fraction of Q-like linear polarization is produced, which is stronger when the los is orthogonal to the Earth magnetic field ($Q\propto(a+b-c)\sin^{2}\theta$). Nevertheless, the ratio of linear to circular polarization intensity is very low ($\approx 10^{-4}$) and the former can be neglected.\cite{hanany}
\section{Signal computation}
Once the ${\mathbf A}$ matrix is computed for a given pressure and temperature configuration at a given altitude, we use the tensor radiative transfer approach for isothermal layers \cite{lenoir67}$^{,}$\cite{lenoir68} to propagate the signal through the atmosphere. According to this theory, the signal emerging from one isothermal layer is a weighted sum of the physical temperatures of the layer itself. The weight function encodes all the information about transfer properties and is a function of the total coherency matrix $\mathbf{A}$.\\*
For the signal computation we used vertical profiles of atmospheric pressure and temperature which are publicly available \cite{spinelli}$^{,}$\cite{tomasi} and the IGRF-2010 model for the Earth magnetic field. The latter describes the geomagnetic field up to an angular scale of $\simeq 15^{\circ}$, corresponding to a typical wavelength of 3000 km along the surface. Smaller scale correction due to magnetized rocks in the crust are not accounted for. 
\section{Results}
\subsection{Frequency templates}
\begin{figure}[!ht]
\centering
\includegraphics[width=.6\textwidth]{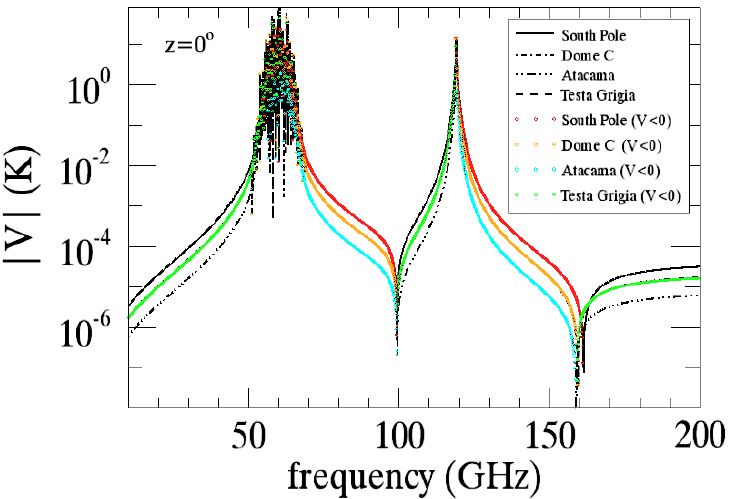}
\caption{Absolute value of polarized atmospheric signal at the zenith for various sites as a function of frequency. Colored lines denote negative values.}
\label{absv}
\end{figure}

We computed the circularly polarized emission at the zenith for various sites suitable for mm astronomy (Chajnantor in Atacama desert, South Pole, Dome C, Testa Grigia) as a function of frequency (fig. \ref{absv}). Differences among those sites are mainly due to the difference of altitudes between them (lower layers of the atmosphere having an higher pressure are in fact the major contributors to the signal) and amplitude and direction of Earth magnetic field, since the signal depends on the scalar product between the latter and the los. We notice that both right handed ($V > 0$) and left handed ($V < 0$) circular polarization are produced and that a sign reversal takes place at frequencies $\nu \simeq 100, 160$ GHz. 
\subsection{Angular templates}
\begin{figure}[!ht]
\centering
\includegraphics[width=.45\textwidth]{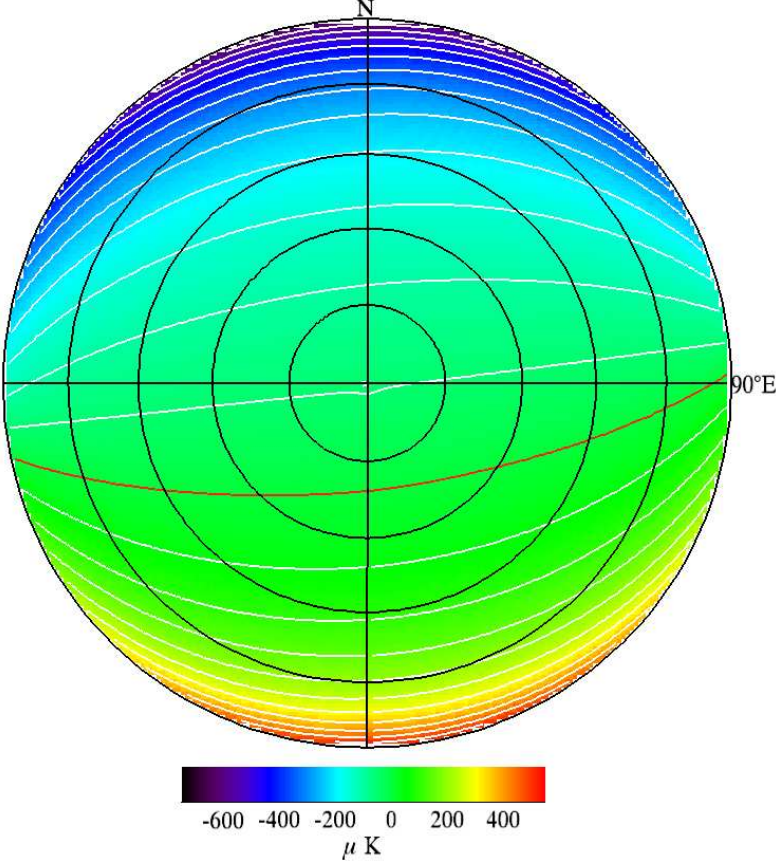}\quad\includegraphics[width=.45\textwidth]{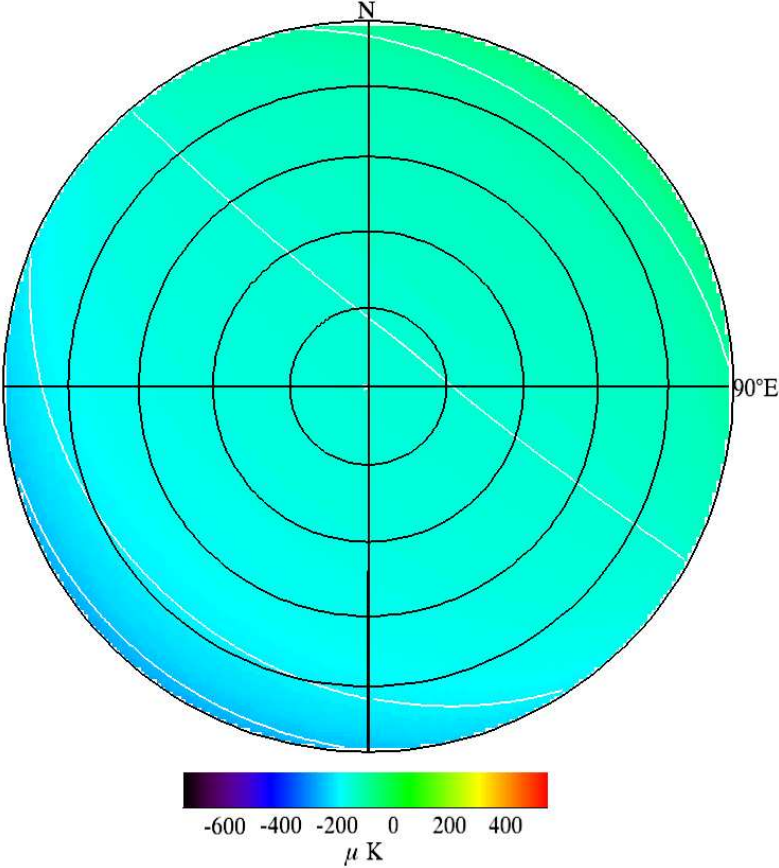}
\caption{Map of atmospheric \o2 circularly polarized signal at $\nu=90$ GHz for Atacama (left) and Dome C (right) in Lambert projection centered on the zenith. Black circles denote elevation intervals of $15^{\circ}$; white (red) contours denote intervals of $50 \mu K$ in brightness temperature (null signal level). }
\label{maps}
\end{figure}

\begin{figure}[!ht]
\centering
\includegraphics[width=.48\textwidth]{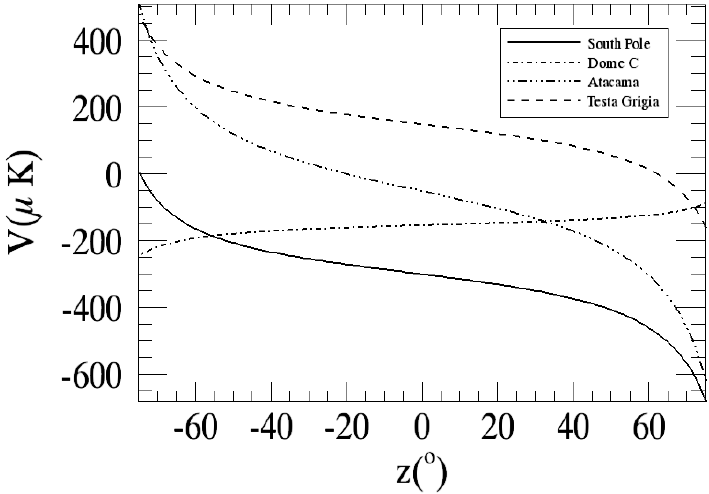}\quad\includegraphics[width=.48\textwidth]{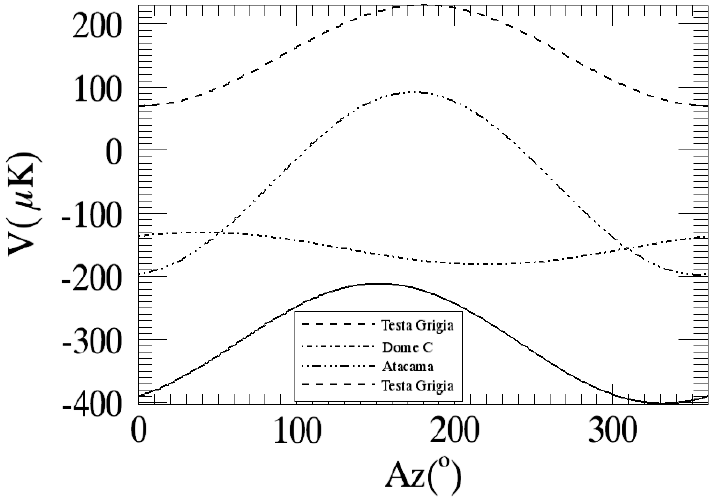}
\caption{ North-South elevation scans (left) and azimuthal scans at constant elevation $el=45^{\circ}$ (right) of the oxygen polarized signal for different sites at $\nu=90$ GHz. }
\label{grad}
\end{figure}

We produced maps of the signal at the chosen sites in local alt-azimuthal coordinates at 90 GHz (see figure \ref{maps} and Spinelli {\it et al.}). Both right handed and left handed circular polarization are produced varying the los.  In some cases a null signal direction is present and corresponds to the angular position where magnetic field and los are orthogonal. As expected, the observed signal depends on a combination of the atmospheric thickness (elevation scans have a zenith secant dependence law) and the magnetic field direction (see fig. \ref{grad}). Such dependencies can lead the two effects to roughly compensate each other producing a nearly constant signal on a large part of the visible sky (see Dome C case in fig. \ref{maps}), or to enhance the signal gradient where the magnetic field is nearly horizontal, as in Atacama case.\\*
Typical signal variation at 90 GHz for North-South elevation scans ranges between $70 \mu K$ (Dome C) and $500 \mu K$ (Atacama) while for a full $360^{\circ}$ azimuthal scan at $el=45^{\circ}$ varies between $50 \mu K$ (Dome C) and $270 \mu K$ (Atacama). 
We also computed the signal gradient for both type of scans in order to give a better description of the signal angular dependence. The maximum value of the gradient takes values in a range  $0.3-5.1 \mu K /^{\circ}$ for elevation scans while for azimuth scans it varies between $0.4-2.5 \mu K/^{\circ}$ with minimum and maximum values at Dome-C and Atacama respectively.\\*
\noindent
Dome C presents the smoother pattern for \o2 polarized signal which thus can be minimized for any kind of differential scanning strategy.  We note however that in case of a realistic application the signal computed in our template has to be integrated with the bandwidth efficiency of the experiment and convolved with its beam profile.

\section{Accuracy}
The accuracy of the 90GHz templates has been estimated taking into account uncertainties and typical variability of the main parameters of the model through Monte Carlo techniques. In particular the influence of uncertainties in the magnetic field model direction and magnitude and secular variation (SV)  have been found negligible for all sites ($\delta V_{rms}<0.2\mu K,\frac{\delta V_{sv}}{\delta t}<0.5\mu K/y$). A violent and rapid event like a solar magnetic storm conversely can affect the accuracy of the template more significantly, causing a variation of $2 - 9 \mu K$ on the level of the V signal. During those events in fact the Earth magnetic field strength can vary as much as $1000$ nT for no more than a few days before quiet conditions are established again.\\*
We then investigated the uncertainty with respect to the accuracy of temperature and pressure profiles using values quoted in literature \cite{tomasi} and found it to be below $2 \mu K$ for all the sites. Day to day variation of temperature and pressure profiles do not affect significantly the signal level while seasonal long term variability of atmospheric parameters affects the accuracy of the template only for polar sites ($3-14 \mu K$), where it concerns the whole air column and not only the lowest layers of the atmosphere like, e.g., the Atacama case.\\*
We note also that the line transition frequencies are known with negligible error bars but the Oxygen absorption parameters, which are crucial to compute the line profiles, have error bars of at least $5\%$ at 90 GHz. This accuracy applies up to $\approx 120$ GHz but at higher frequencies, where models and values of Oxygen absorption coefficients are not yet validated, results should be used with caution.

\end{document}